\documentstyle[aps,prb,twocolumn,psfig]{revtex}

\newcommand{\V}{NaV$_2$O$_5$~}
\newcommand{\nn}{\nonumber}
\newcommand{\ra}{\rangle}
\newcommand{\la}{\langle}
\newcommand{\vT}{\bf T\rm}

\begin{document}

\draft

\title{Isospin model for suppression of charge order in hole- doped \V}

\author{Peter Thalmeier}

\address{Max-Planck-Institute for the Chemical Physics of Solids,
N\"othnitzer Str.40, 01187 Dresden, Germany}

\maketitle

\begin{abstract}
The layered perovskite compound \V exhibits charge ordering (CO) and spin
gap formation which is driven by inter- site Coulomb
forces and exchange dimerisation. Experimentally it was found that the
CO transition is rapidly reduced by doping. Charge
ordering in \V has a quasi 1D character due to inter- ladder frustration
in the Trellis lattice. Based on an extended pseudospin
(T=1) approach for hole doped \V ladders we construct an Ising like
model that explains qualitatively the destruction of the CO phase as
function of hole doping $\delta$ and show how the CO parameter depends on
$\delta$ and the ratio of Coulomb and hopping energies. 
\end{abstract}



\section{Introduction}
\label{intro}

The layered perovskite \V shows an interesting interplay between charge
ordering (CO) of 3d- electrons on the rungs of V-V ladders and a spin-
Peierls transition associated with spin gap formation first observed
in \cite{Isobe96} below T$_c$= 34 K. This transition is associated
with a distortion of the high temperature orthorhombic structure of
\V. The latter can be viewed as Trellis lattice with
ab- planes stacked along c. They consist of ladders with
V-O-V rungs (inset of Fig.~\ref{ENERGY}) shifted alternatingly along b by
half a lattice constant. The precise low temperature structure is still
controversial\cite{Luedecke99,deBoer00,Sawa02,Bernert01,Chatterji02} with
the proposed models differing in the number of inequivalent V and Na
sites and an associated charge ordering on each or only on every second
ladder. The slow progress and ambiguity of experimental results has
hindered theoretical understanding considerably. However it has become
clear now that there is a close connection between charge ordering
on the V-V rungs and exchange dimerisation on the
ladders\cite{Bernert02} which leads
to the spin gap formation directly seen in inelastic neutron
scattering\cite{Grenier02}. It was realized early on\cite{Thalmeier98}
that the CO on
the V- rungs may be described within an Ising model in a transverse
field (ITP) where the isospin (S=$\frac{1}{2}$) represents the charge degree
of freedom on the V-V rung with inter- site Coulomb forces correspond to
Ising interactions and the intra-rung hopping to a transverse
field. This is due to the charge transfer (CT) gap of $\sim$ 0.4-0.5 eV
\cite{Yaresko00,Cuoco99} which leads to an insulating state of \V 
despite having only one 3d- electron per rung (quarter-filled
ladders). Therefore only singly occupied rungs have to be considered
which may be represented by an isospin and the real spin. A detailed
model of the combined Ising- spin Peierls system and its phase
transitions has been discussed in \cite{Bernert02}.

The charge ordering in a Trellis lattice compound is severely
hindered by frustration of inter-site Coulomb forces which is directly
obvious in the Ising model representation. In fact on a rigid lattice
CO would be prohibited since the lattice of V-V rungs corresponds to a
trigonal covering lattice of Ising spins. The simultaneous structural
distortion at T$_c$  reduces the frustration by leading to
inequivalent adjacent ladders. The driving mechanism for the
transition are inter- site Coulomb interactions and
exchange-striction. The gain in Coulomb energy (V$_b$) by complete CO
in a rung is in competition with the loss of intra-rung kinetic energy
(t$_a$), the control parameter for CO is therefore $\lambda$
=$2t_a/V_b$ (for convenience the inverse definition of \cite{Yushankhai01} is
used here). The very low CO transition temperature suggest that in \V
one is close to the quantum critical point $\lambda_c$=1 of the
corresponding ITP model\cite{Yushankhai01}. The charge ordering is
then essentially of 1D Ising type, this means one has long range
correlations along
the ladder and no correlations between them, this was indeed obtained
 in exact diagonalisation studies of the ITP\cite{Langari01}. The
ordering may then be viewed as arising  from Ising like correlations
along the chain which lead to a true 2D broken symmetry state at finite T
due to a staggered longitudinal Ising field set up by the distortion
of the neighboring ladders \cite{Yushankhai01,Bernert02}.

In this context it is suggestive that even small perturbations of the
1D Ising correlations might reduce or destroy the CO state in
\V. This can be achieved by reducing the filling factor n of the 1D
ladders below $\frac{1}{4}$ by doping with holes. This introduces
"empty" rungs into the ladder which cut the Ising bonds. The ensuing
destruction of long range 1D correlations should then strongly reduce
T$_c$ as function of the hole concentration $\delta$
(n=$\frac{1}{4}$-$\delta$). This has indeed been found by introducing
holes in the ladders through Na- deficiency doping \cite{Isobe97} where
a few per cent holes are sufficient to destroy the CO state and the
associated spin Peierls transition. Rapid suppression of charge order has
also been found in various other doping series, i.e. replacing Na by
Li and K (isoelectronic) or Ca (electron-doping).
\cite{Isobe98,Dischner01,Lohmann00}
This may be observed directly by specific
heat measurements\cite{Dischner01} which show a progressive
suppression of $\Delta$C(T$_c$) with increasing doping.
It is also seen in the susceptibility\cite{Isobe97,Isobe98}
which shows a closing of the CO induced spin gap. 

Most importantly in this doping range \V remains an insulator which is
not easy to understand within a Hubbard like model for the quarter
filled ladder\cite{Bernert01,Bernert02}, possibly 1D loalization
effects play a role, indeed the conductivity was found to exhibit
variable range hopping behaviour\cite{Isobe97} for hole doping. The
notion that charge ordering T$_c$ is reduced because doping creates
empty rungs which cut the Ising
bonds along the ladder is however completely static and invokes a
disordered state. It lacks the important ingredient of kinetic energy
gain due to the hopping of holes between adjacent rungs leading to a
homogeneous state. This problem has sofar not been considered theoretically. 

In this work we will give a more solid theoretical basis for the
influence of hole doping on CO in \V. This will again be done in the context
of an effective isospin approach. In sect.~\ref{elstruc} we give the
basics of the electronic structure of \V and discuss the isospin model
for the undoped case in sect.~\ref{undoped}. Subsequently in
sect.~\ref{doped} we show that even in the doped case
the low energy Hamiltonian may be projected to an effective Ising type
form, albeit with an Ising spin T=1 to accomodate the holes on empty
rungs. In sect.~\ref{co} we discuss how order parameter, kinetic energy
and chemical potential (doping level) are represented in the isospin
formalism and their respective coupled mean field equations in the
ground state in sect.~\ref{mf}. They are solved in sect.~\ref{numerical}
numerically and we discuss the variation of the CO parameter with the
control parameter $\lambda$ and the hole doping $\delta$ which leads to the
zero temperature CO phase diagram. Finally sect.~\ref{summary} gives
summary and outlook.

\section{Electronic structure of \V}
\label{elstruc}

The electronic structure of the charge transfer insulator \V has been
investigated\cite{Smolinski98,Yaresko00} within the LDA and LDA+U
approach respectively. A mapping of the resulting band structure to
an extended
tight binding (TB) model involving both V 3d and O 2p orbitals has also
been performed\cite{Yaresko00}  and an excellent fitting of the LDA+U
bandstructure was obtained. For our present purpose it is sufficient
to use a reduced
TB model involving only V 3d$_{xy}$ orbitals. The hopping elements
involved in this model are defined in Fig.~\ref{ENERGY}. There are four V
atoms in the unit cell resulting in two 3d$_{xy}$ occupied valence and
two empty 3d$_{xy}$ conduction bands separated by a charge transfer gap of
about 0.3 meV which is largely determined by the effective intra- rung
hopping t$_a$. In the effective TB model this gap may be
interpreted as bonding- antibonding gap of V 3d$_{xy}$ orbitals of a
rung. Furthermore the valence and conduction band are each almost
twofold degenerate due to the weak hopping between
ladders. The 3d$_{xy}$ bonding bands have a considerable dispersion of
order t$_b$+t$_d$ along the b- direction while the dispersion
t$_b$-t$_d$ of 3d$_{xy}$ antibonding bands is almost zero due to
accidental near equality t$_b\simeq t_d$.

\section{The isospin model Hamiltonian for undoped \V}
\label{undoped}

The basic many body Hamiltonian describing \V is of the single band
(3d$_{xy}$) extended Hubbard type \cite{Thalmeier98,Bernert02}. As
shown in \cite{Thalmeier98}, due to the presence of a CT gap the
charge degrees of freedom may be represented by an effective ITP
Hamiltonian (T=$\frac{1}{2}$) which is obtained by projecting the
original Hamiltonian to the subspace of singly occupied
rungs\cite{Bernert02}. The states $|-\frac{1}{2}\ra$,
$|\frac{1}{2}\ra$ of the T=$\frac{1}{2}$ isospin describes occupied
d$_{xy}$- orbitals on the left and right V atom of each rung
respectively. A simplified version of the resulting ITP Hamiltonian
for a single ladder is given by\cite{Yushankhai01}

\begin{eqnarray}
\label{HamITP} 
H_{ITP}&=&2V_b\sum_{\la ij\ra}[T_{zi}T_{zj}+\frac{1}{4}] -2t_a\sum_iT_{ix}\nn\\
&&-h_s^0\sum_i(-1)^iT_{zi}
\end{eqnarray}
 
Here t$_a$ is the hopping across the rung and V$_b$ the Coulomb
interaction along the leg of a ladder. The
last term is due to a staggered longitudinal isospin field
h$_{si}$ =(-1)$^i$h$_s^0$ which simulates the influence of the neighboring
ladders\cite{Yushankhai01}. For $\lambda$ = 2t$_a$/V$_b<\lambda_c$=1
one obtains an AF ordering of isospins which
corresponds to the zig-zag ordering of 3d- electrons on the ladder as
illustrated in Fig.~\ref{ENERGY}. It was shown in \cite{Bernert02} that the
interaction parameters in Eq.~\ref{HamITP} are renormalized by virtual
processes via doubly occupied rungs or sites which depends on the spin
state. This leads to a coupling of isospin (charge) and spin degrees
of freedom whose contribution is neglected here.

The zig-zag CO in a ladder is characterized by the order parameter 
$\la T_{zi}\ra=(-1)^ix$ at zero temperature with ($\lambda\leq$1)

\begin{equation} 
x= \left[1-\left(\frac{\lambda}{\lambda_c}\right)^2\right]^\beta
\label{OP}
\end{equation}

The exact solution of the 1D ITF model has the exponent
$\beta=\frac{1}{8}$ and $\lambda_c$=1. It is understood that an
infinitesimal staggered field h$_{si}$ has to be applied to lift the
twofold degeneracy of the ordered ground state. A truncated mean field (mf)
approximation suppressing double counting of bonds in the molecular
field leads to a critical value $\lambda_c$=1. This is the same 
quantum critical point that separates the CO ($\lambda <$1) phase
from the disordered phase  ($\lambda >$1) as for the exact
solution\cite{Yushankhai01}. Naturally the mf critical OP exponent 
$\beta=\frac{1}{2}$ is different.

\section{T=1 isospin model for doped \V}
\label{doped}

The charge ordering described by the ITP model and its implications
for the (spin-) superexchange have been described in detail in
\cite{Yushankhai01,Bernert02}. In the present work we want to
investigate whether the convenient and physically appealing isospin
approach for CO in \V can also be applied to the hole doped compound
to study the suppression of CO. For
this purpose the hole states, that is empty V-V rungs with zero 3d-
occupation, also have to be represented by an isospin degree of
freedom. As in the stoichiometric case\cite{Yushankhai01,Bernert02}
doubly occupied rung states either have a large charge transfer energy
(t$_a$) or a large on-site Coulomb repulsion and therefore contribute
only as intermediate states by renormalizing the interaction
parameters of the effective Hamiltonian. Thus the model has to
incorporate three basis states, two singly occupied and one empty rung state
which suggests to use a T=1 isospin instead of T=$\frac{1}{2}$ for the
stoichiometric ($\delta$=0) case. The mapping of electronic states to isospin
states for rung i is then defined as

\begin{eqnarray}
|T_z=-1\ra_i &=&|10\ra_i = c^\dagger_{Li}|00\ra_i\nn\\
|T_z= 0\ra_i &=&|00\ra_i\\
|T_z= 1\ra_i &=&|01\ra_i = c^\dagger_{Ri}|00\ra_i\nn
\end{eqnarray}

\begin{figure}
\centerline{\psfig{figure=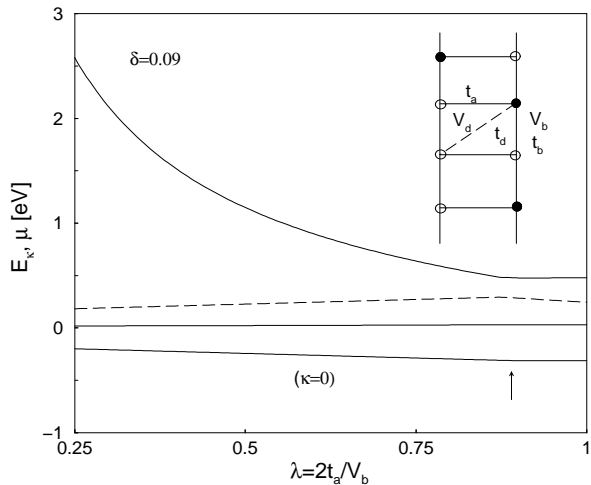,width=0.9\columnwidth}}
\vspace{0.5cm}
\caption{Mean field eigenvalues (full lines) E$_\kappa$ ($\kappa$=
0-2) of Eq.~(\ref{MFAB}) and 'chemical potential' $\mu$ (dashed line) as
function of the control parameter $\lambda$. Arrow indicates CO
transition. Inset shows the \V ladder along b with effective V-V rungs
along a. Closed and open circles correspond to occupied and empty
d$_{xy}$- orbitals. Hole doped rung has two empty d$_{xy}$- orbitals.
Only the ground state ($\kappa$=0) is occupied for each rung with a hole
count given by Eq.~\ref{HOLEX}.
Hopping and interaction parameters are indicated. We used t$_a$=
0.380 eV, t$_b$= t$_d$ =0.170 eV; for numerical reasons the latter is
twice the value used in \cite{Yushankhai01}.}
\label{ENERGY}
\end{figure}

with $|00\ra_i$ denoting the empty rung state. We first express the
inter- rung Coulomb energy already present in Eq.~(\ref{HamITP}) in
the above extended isospin
basis. The projectors to this basis may be written in terms of the T=1
isospin operators \vT= (T$_x$,T$_y$,T$_z$) as 

\begin{eqnarray}
P_L&=&\frac{1}{2}T_z(T_z+1)\nn\\
P_R&=&\frac{1}{2}T_z(T_z-1)\\
P_0&=&1-T_z^2\nn
\end{eqnarray}

where P$_0$+P$_1$=1 and P$_1$=P$_L$+P$_R$ holds. The other projector
properties P$_0$P$_1$= P$_1$P$_0$=0 and P$^2_\lambda$= P$_\lambda$
($\lambda$=0,L,R) are easily verified. Using these projection operators
the intra- ladder Coulomb interactions V$_b$ between adjacent rungs
i,j may be represented by the Hamiltonian

\begin{eqnarray}
H_C(i,j)=\frac{1}{2}(V_b+V_d)T^2_{zi}T^2_{zj} + 
\frac{1}{2}(V_b-V_d)T_{zi}T_{zj}
\end{eqnarray}

Compared to the T=$\frac{1}{2}$ case of Eq.~(\ref{HamITP}) H$_C$ also
contains now biquadratic terms in the isospin. Similarly we construct the
intra- rung kinetic energy responsible for the charge transfer
gap. Since the R,L- occupation of each rung is now described by
$|T_z=\pm 1\ra$ respectively the intra-rung hopping term must change
the isospin by $\pm$2 units. The corresponding rung Hamiltonian then
reads

\begin{eqnarray}
H_K(i)=-\frac{1}{2}t_a(T^2_{+i}+T^2_{-i})= 
-t_a(T^2_{xi}-T^2_{yi})
\end{eqnarray}

which should be compared to the second term in Eq.~(\ref{HamITP}) for
T=$\frac{1}{2}$. Finally there has to be an additional term describing
the hole kinetic energy, i.e. when the empty rung (Fig.\ref{ENERGY}) is
shifted along the ladder direction. Such processes are
described by inter- rung terms like 

\begin{eqnarray}
H'_K(i,j)&=&2t_+[O_{zy}(i)O^\dagger_{zy}(j)+O^\dagger_{zy}(i)O_{zy}(j)]\nn\\
&+&2t_-[O_{zx}(i)O^\dagger_{zx}(j)+O^\dagger_{zx}(i)O_{zx}(j)]
\end{eqnarray}

with i,j denoting rungs along the ladder. Here
t$_\pm$=$\frac{1}{2}$(t$_b\pm$t$_d$) and  O$_{zy}$=iT$_z$T$_y$ and
O$_{zx}$=T$_z$T$_x$ may be interpreted as a quadrupolar operator in
the T=1 isospin space. In the following we consider only the special
case t$_b\simeq$ t$_d$ which is appropriate for \V
\cite{Yaresko00}. Then the second term above vanishes and the total
T=1 model Hamiltonian for the doped ladder now reads

\begin{eqnarray}
\label{HAM1}
H&=&\sum_i[\epsilon T_{zi}^2+t_a(T_{xi}^2-T_{yi}^2)-h^0_{si}T_{zi}]\nn\\
&&+2t_b\sum_{\la ij\ra}[O^i_{zy}O^{j\dagger}_{zy}+O^{i\dagger}_{zy}O^j_{zy}]\\
&&+V_-\sum_{\la ij\ra}T_{zi}T_{zj}+V_+\sum_{\la ij\ra}T^2_{zi}T^2_{zj}\nn
\end{eqnarray}

Here $\epsilon$ is the on-site orbital energy and h$^0_s$(i)=
h$^0_s(-1)^i$ is again a staggered field due to the influence of
neighboring ladders, $\la ij\ra$ denotes n.n. rungs along the ladder.
Furthermore we have defined
V$_\pm=\frac{1}{2}(V_b\pm V_d)$. If we neglect the ladder diagonal
Coulomb interaction (V$_d\equiv$0) then V$_\pm$=$\frac{1}{2}$V$_b$ 
 leading to a single control parameter $\lambda$= 2t$_a$/V$_b$.
Although we expect the isospin model a convenient framework to discuss
the energetics of charge ordering as function of doping it can however
not give answer to the question why the doped system stays an
insulator, for this problem one also has to include the doubly
occupied rung states. It was argued in \cite{Bernert01,Bernert02} that
even under a finite charge transfer between ladders, i.e. deviations from
quarter-filling, an insulating state can be expected for the relevant
hopping and interaction parameters realised in \V.

\begin{figure}
\centerline{\psfig{figure=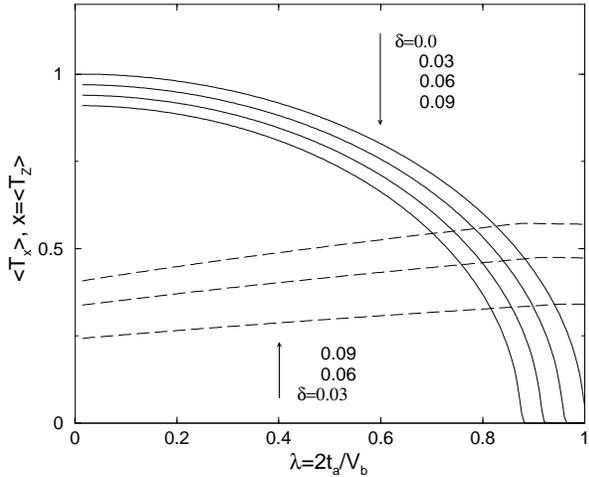,width=0.9\columnwidth}}
\vspace{0.5cm}
\caption{zig-zag charge order parameter $\la T_z\ra$ as function of 
control parameter $\lambda$ for small and moderate hole doping
$\delta$ (full lines). Dashed lines represent average hole hopping
energy $\la T_x\ra$ (units: t$_b$), for $\delta$=0: $\la T_x\ra\equiv$ 0}
\label{ORDER}
\end{figure}

\section{Charge order parameter, doping and 'chemical potential'}
\label{co}

It is obvious that the model defined by eq.~\ref{HAM1} is much more
complex due to the additional hole state (T$_z$=0) and has no exact 1D
solution like the T=$\frac{1}{2}$ ITF model. As mentioned previously
the truncated mf approximation for the charge order parameter of the
ITF model has the same quantum critical point $\lambda_c$=1 as the exact
solution. It is therefore plausible that a mf treatment of 
Eq.~\ref{HAM1} in the doped case still gives a qualitatively correct
picture for the conditions of charge ordering. As in the ITF case the
order parameter for V$_b>$V$_d$ should be staggered, i.e zig-zag CO
should be realized, then for the two sublattices we have

\begin{eqnarray}
x= \la T_{zA}\ra = - \la T_{zB}\ra
\end{eqnarray}

Now x not only depends on the control parameter $\lambda$=2t$_a$/V$_b$
but also on the number of holes or unoccupied
rungs. The representation of the hole number operator in fermionic and isospin
language is given by

\begin{eqnarray}
\hat{\delta}_i&=&1-\hat{n}_i=1- c^\dagger_{Li}c_{Li}
-c^\dagger_{Ri}c_{Ri}\nn\\
\hat{\delta}_i&=&1-P_{iL}-P_{iR}= 1-T^2_{zi}
\end{eqnarray}

so that the hole concentration is obtained as 

\begin{eqnarray}
\delta=1-\la T^2_z\ra
\end{eqnarray}
 
Since it is externally fixed by the level of doping the local orbital
energy has to be adjusted selfconsistently
for each parameter set such that $\delta$ is obtained. This can be done by
adding a 'chemical potential' , i.e. replacing
$\epsilon\rightarrow\epsilon-\mu$. For each microscopic parameter set and order
parameter x, the chemical potential $\mu$ has to be adjusted
appropriately to obtain the physical hole number.
 
\begin{figure}
\centerline{\psfig{figure=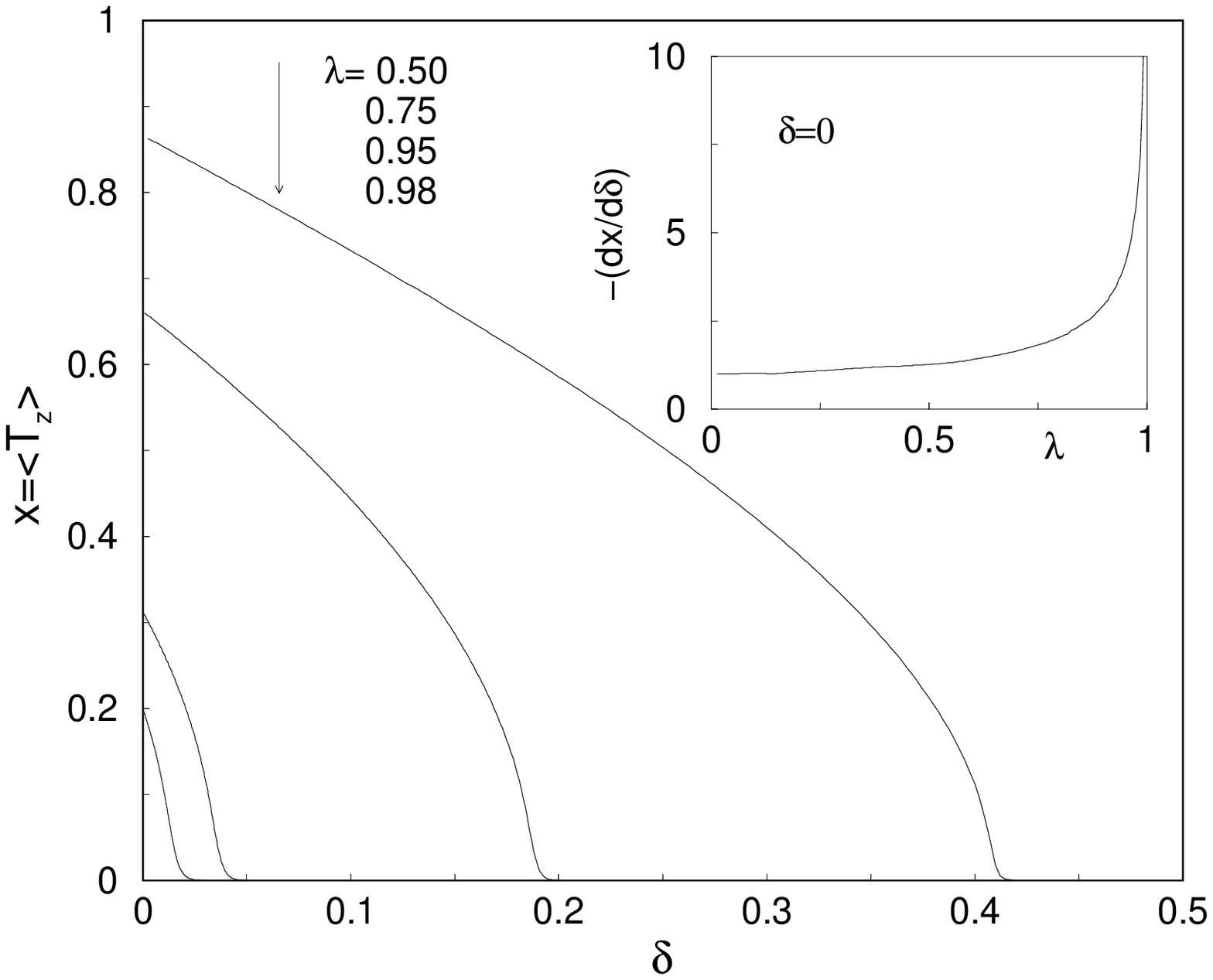,width=0.9\columnwidth}}
\centerline{\psfig{figure=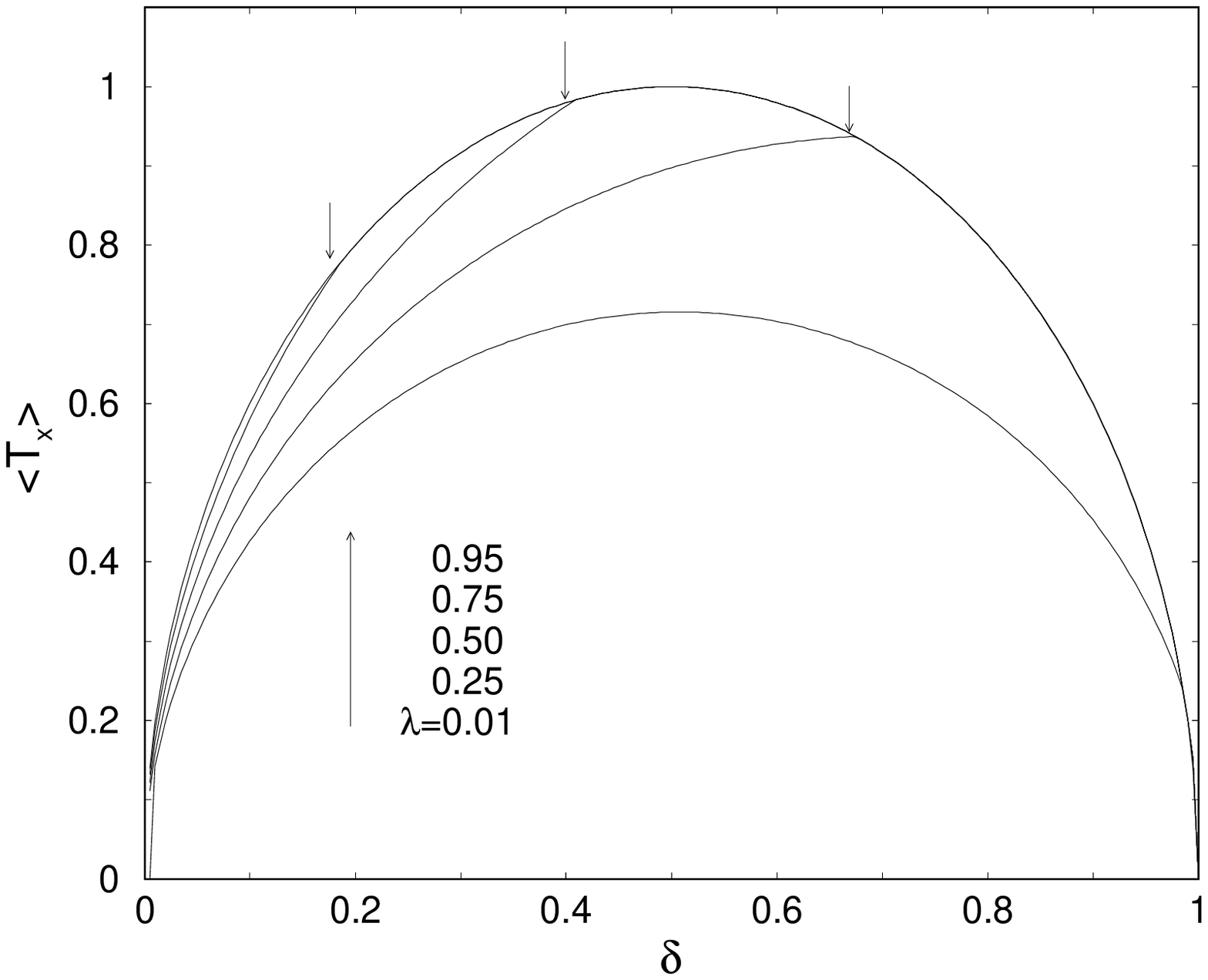,width=0.9\columnwidth}}
\vspace{0.5cm}
\caption{upper panel: melting of CO with hole doping $\delta$ for
various $\lambda$ given in decreasing order. The absolute slope value
-(dx/d$\delta$) increases strongly when approaching the quantum critical
point $\lambda_c$=1 of CO; this is shown in the inset.
lower panel: Hole kinetic energy $\la T_x\ra$ (in units of t$_b$) as
function of $\delta$ in the complete doping range, plotted for five
values of $\lambda$ corresponding to curves in increasing
order. Downward arrows indicate CO transition (x=0) where the curves
for different $\lambda$ merge.}
\label{DOPING}
\end{figure}

\section{Mean field approximation and solution for CO}
\label{mf}

The usefulness of a simple mf solution relies on the Ising type nature
of the present model. In this approximation the doping term $\sim t_b$
in Eq.~\ref{HAM1} simplifies considerably using O$_{zy}$+O$^\dagger_{zy}$ =
T$_x$ and O$_{zy}$-O$^\dagger_{zy}$=iT$_y$, we obtain

\begin{eqnarray}
H'_{Kmf}(i)&=& -2t_b[O_{zy}\la O_{yz}\ra+O_{yz}\la O_{zy}\ra]\nn\\
&=& -t_b(\la T_x\ra T_x(i)+\la T_y\ra T_y(i))\\
&=&-t_b\la T\ra (\cos\phi T_x+\sin\phi T_y)\nn
\end{eqnarray}

where $\la T\ra$=$\sqrt{\la T_x\ra^2+\la T_y\ra^2}$ and $\phi$ which is
given by $\tan^{-1}\phi$=$\la T_y\ra$/$\la T_x\ra$ is an arbitrary
angle which may be chosen as $\phi$=0. Then only $\la T_x\ra$ appears and
the quantity W= t$_b\la T_x\ra$
may be interpreted as the average inter- rung hopping amplitude of the
hole. Together with the remaining terms in Eq.~\ref{HAM1} we get the
following rung mf Hamiltonian for A,B sublattices: 

\begin{eqnarray}
H^{mf}_A&=&[\epsilon +V_+(1-\delta)-\mu]T_{zA}^2-t_a(T^2_{xA}-T^2_{yA})\nn\\
&&-(h^0_s+V_-x)T_{zA} -WT_{xA}\nn\\
H^{mf}_B&=&[\epsilon +V_+(1-\delta)-\mu]T_{zB}^2-t_a(T^2_{xB}-T^2_{yB})\\
&&+(h^0_s+V_-x)T_{zB} -WT_{xB}\nn
\end{eqnarray}

The effective mf orbital energies and staggered fields are then given
by 

\begin{eqnarray}
\hat{\epsilon}=\epsilon-\mu +V_+(1-\delta)\nn\\
h_s=h^0_s+V_-x\nn
\end{eqnarray}

In the T=1 isospin space the mf Hamiltonian for A,B sublattices can
explicitly be written as

\begin{eqnarray}
\label{MFAB}
H^{mf}_{A,B}=
&&\left(\matrix{
\hat{\epsilon}\mp h_s & -W/\sqrt{2} & -t_a                   &\cr
-W/\sqrt{2}           & 0           & -W/\sqrt{2}            &\cr
-t_a                  & -W/\sqrt{2} & \hat{\epsilon}\pm h_s  &\cr
}\right)
\end{eqnarray}

where A,B corresponds to upper and lower signs respectively. These
equations have to be selfconsistently solved for the charge order
parameter x, keeping the hole concentration fixed by varying the
chemical potential $\mu$. Only the T=0 case is considered, then the
expectation values $\la X\ra =\la
\psi^{mf}_0|X|\psi^{mf}_0\ra$ are obtained from the ground state
$\psi^{mf}_0$ solution of Eq.~\ref{MFAB}. For the actual calculation,
in addition to assuming t$_a\simeq$ t$_d$ which has been justified in
 \cite{Yaresko00}, we take for simplicity $V_d\simeq$ 0. If we
suppress the hole
hopping described by the matrix elements $\sim$W for the moment, the
T$_z$=0 and T$_z=\pm 1$ subspaces disconnect and  
the mf ground state and its energy of the isolated rungs can be written
explicitly as\cite{Yushankhai01}

\begin{eqnarray}
E_0&=&\hat\epsilon-\frac{1}{2}
\sqrt{(\Delta\epsilon)^2+ (2t_a)^2}\nn\\
|\psi^{mf}_0\ra_i&=&u_{1i}|1\ra+u_{2i}|-1\ra
\end{eqnarray}

where $\Delta\epsilon$=V$_b$x is the splitting of L,R orbital
energies and the staggered mf coefficients are given by 

\begin{equation} 
u_{1,2i}=\sqrt{\frac{1}{2}\left[1\pm\frac{(-1)^i\Delta\epsilon_i}
{\sqrt{(\Delta\epsilon)^2 +(2t_a)^2}}\right]}
\end{equation}

When the effect of hole hopping along the legs (W$>0$) is now
included the mf eigenstates will be mixtures of the isolated rung
states $|\psi^{mf}_0\ra_i$ and the hole state $|0\ra_i$. The hole
concentration is then given by 

\begin{eqnarray}
\label{HOLEX}
\delta&=&1-\la\psi^{mf}_0|T^2_z|\psi^{mf}_0\ra
\end{eqnarray}

The new mf- ground state energy, its wave function and associatied
expectation values of T$_x$, T$_z$ and T$^2_z$ then have to be
calculated selfconsistently from Eq.~\ref{MFAB}. The numerical results
will be discussed in the next section.

\section{Numerical solution, doping dependence of CO and phase diagram} 
\label{numerical}

In this section the numerical results for the above model will be
presented. The main goal is the determination of the charge order
parameter x($\delta,\lambda$) as function of the hole concentration
$\delta$ and the control parameter $\lambda$= 2t$_a$/V$_b$. To achieve
a given hole number $\delta$= 1-$\la T_z^2\ra$ the 'chemical potential'
$\mu$ has to be adjusted selfconsistently together with the
determination of the order parameter x=$\la T_z\ra$ and the normalized
kinetic energy $\la T_x\ra$=W/t$_b$. Physically $\mu$ corresponds to an orbital
energy shift of the empty and singly occupied rung states. This energy
shift together with the mf eigenvalues of Eq.~\ref{MFAB} as function
of the control parameter $\lambda$ is shown in Fig.~\ref{ENERGY} for
$\delta$=0.09. 

\begin{figure}
\centerline{\psfig{figure=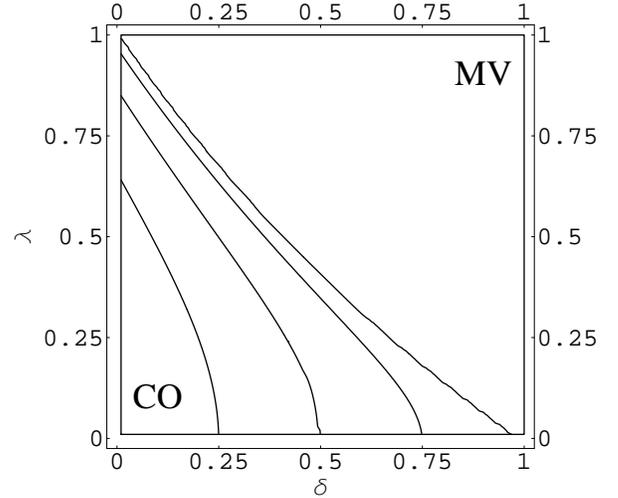,width=0.9\columnwidth}}
\vspace{0.5cm}
\caption{CO/MV phase diagram in the $\delta,\lambda$- plane. Contour
lines for CO phase with $\delta$=$\la T_z\ra$ for x=0.01, 0.25, 0.5, 0.75
(moving to lower left corner) are also shown. MV denotes the
disordered homogeneus mixed valent phase.}
\label{PHASE}
\end{figure}

The corresponding variation of the charge order parameter x for
several hole concentrations is shown in Fig.~\ref{ORDER}. As the hole
concentration increases our results show the expected qualitative
behaviour: i) the CO even for large V$_b$ (small $\lambda$) is no
longer complete due to the presence of empty rung states and its variation
with $\lambda$ becomes essentially flat. ii) the critical value
$\lambda_c$ for the CO transition to appear is reduced
from the value $\lambda_c$=1 of the undoped ITF model. In addition the
kinetic energy $\la T_x\ra$ is shown. It increases
almost linearly with $\lambda$= 2t$_a$/V$_b$ and rapidly with the
hole concentration. 

The complementary reduction of charge order as function of the hole
concentration for constant $\lambda$ is shown in Fig.~\ref{DOPING}. Most
significantly for small $\delta$ one observes a rapid reduction of the
CO- parameter with increasing hole concentration. The slope of this
reduction increases even further when $\lambda$ approaches the quantum
critical point  $\lambda_c$=1 of CO from below. This is shown in more
detail in the inset of Fig.~\ref{DOPING}. The origin of this drastic
reduction of CO for small $\delta$ lies in the rapid increase of hole
kinetic energy with doping as can bee seen in the lower panel of
Fig.~\ref{DOPING}.  

Our results shown in Fig.~\ref{DOPING} describe qualitatively
correct the experimental situation, namely a destruction of the CO
state in \V by a few percent hole doping \cite{Dischner01}. For
example assuming a value $\lambda$ =0.95 close to $\lambda_c$ =1
one obtains a critical
concentration $\delta$ = 0.03. The maximum size of CO at $\delta$=0
for this $\lambda$- value is already considerably reduced to
x$_0\simeq$ 0.3. The experimental value of the nominal hole concentration for
suppression of CO in Na- deficient \V is indeed about $\delta$ =
0.03\cite{Isobe98}.

Finally we summarize the results of our calculations in Fig.~\ref{PHASE}
which shows the phase diagram of CO in the ($\delta$,$\lambda$)-
plane and the size of the charge order parameter x as a contour plot
in the ordered regime. 

\section{Summary and Oulook}
\label{summary}

In this work we have proposed an extended isospin approach proposed
in \cite{Thalmeier98} to the charge ordering transition in Na-
deficient (hole-doped) \V. The additional hole (empty rung) state can
be acommodated within a T=1 isospin as the T$_z$=0 component wheras
the (left and right) occupied rung states are represented by the
T$_z=\pm$1 components. The effective isospin Hamiltonian for the doped
system is of more complex type than the simple T=$\frac{1}{2}$ ITF
model in the undoped case due to the existence of biquadratic terms.
Contrary to the ITF model there is no exact solution. However it was
argued that a mf treatment of the extended model for the doped
ladder still gives a qualitatively correct description of charge order
and its destruction by doping.

Indeed it was found that the CO ground state is rapidly destroyed by
hole doping and the size of the slope of the order parameter reduction
-dx/d$\delta$ increases strongly when one approaches the quantum
critical point ($\lambda\leq$ 1) which corresponds qualitatively to
the observed behaviour in \V. A phase diagram in the doping($\delta$) 
- control parameter ($\lambda$) plane was derived showing the evolution of the
charge order parameter.

Naturally the present model should be investigated beyond the present
mf approximation. As shown in the case of the ITF model for pure
\V\cite{Langari01} the exact diagonalisation Lanczos approach of
finite ladders or clusters with the Trellis lattice structure is a
powerful method to gain more insight into the appearance of charge
order. It can also be used for the present extended T=1 isospin model
of hole- doped \V.

\vspace{1cm}
\noindent
{\em Acknowledgement}\\ 
The author would like to thank A. Bernert, Y. Jang, A. Yaresko and
V. Yushankhai for useful discussions and remarks.



\end{document}